\documentclass[conference]{IEEEtran}
\usepackage[letterpaper, left=0.65in, right=0.65in, bottom=1in, top=0.8in]{geometry}
\IEEEoverridecommandlockouts

\usepackage[acronym]{glossaries}
\newacronym{WMMSE}{WMMSE}{weighted minimum mean square error}
\newacronym{FCL}{FCL}{fully connected layer}
\newacronym{NLOS}{NLOS}{non-line-of-sight}
\newacronym{LOS}{LOS}{line-of-sight}
\newacronym{MAC}{MAC}{multiply-accumulate}
\newacronym{BS}{BS}{base station}
\newacronym{PS}{PS}{phase-shifter}
\newacronym{NAS}{NAS}{neural architecture search}
\newacronym{PTQ}{PTQ}{post-training quantization}
\newacronym{QAT}{QAT}{quantization-aware training}
\newacronym{LSQ}{LSQ}{Learned Step Size Quantization}
\newacronym{STE}{STE}{Straight Through Estimator}
\newacronym{MPQ}{MPQ}{Mixed-Precision Quantization}
\newacronym{RL}{RL}{reinforcement learning}
\newacronym{AP}{AP}{analog precoder}
\newacronym{FC-HBF}{FC-HBF}{fully-connected HBF}
\newacronym{FSA-HBF}{FSA-HBF}{fixed subarray HBF}
\newacronym{DSA-HBF}{DSA-HBF}{dynamic subarray HBF}
\newacronym{BF}{BF}{beamforming}
\newacronym{UE}{UE}{user equipment}
\newacronym{AWGN}{AWGN}{additive white gaussian noise}
\newacronym{MIMO}{MIMO}{multiple-input multiple-output}
\newacronym{MISO}{MISO}{multiple-input single-output}
\newacronym{RF}{RF}{radio frequency}
\newacronym{RIS}{RIS}{reconfigurable intelligent surfaces}
\newacronym{IOT}{IOT}{internet-of-things}
\newacronym{QNN}{QNN}{Quantized Neural Network}
\newacronym{CL}{CL}{convolutional layer}
\newacronym{UdeM-NLOS}{UdeM-NLOS}{Universite De Montreal-Non-Line-Of-Sight}
\newacronym{Okapark-LOS}{Okapark-LOS}{Okapark-Line-Of-Sight}
\newacronym{FDD}{FDD}{frequency division duplex}
\newacronym{TDD}{TDD}{time division duplex}
\newacronym{CSI}{CSI}{channel state information}
\newacronym{DNN}{DNN}{Deep Neural Network}
\newacronym{DP}{DP}{digital precoder}
\newacronym{DL}{DL}{deep learning}
\newacronym{SVD}{SVD}{singular-value decomposition}
\newacronym{CNN}{CNN}{convolution neural network}
\newacronym{FDP}{FDP}{fully digital precoder}
\newacronym{SE}{SE}{spectral efficiency}
\newacronym{OFDM}{OFDM}{orthogonal frequency division multiplexing}
\newacronym{OMP}{OMP}{orthogonal matching pursuit}
\newacronym{FL}{FL}{fully-connected layer}
\newacronym{HSHO}{HSHO}{Hybrid Structured Heuristic Optimization}
\newacronym{HBF}{HBF}{hybrid beamforming}
\newacronym{IA}{IA}{initial access}
\newacronym{mm-Wave}{mm-Wave}{millimeter wave}
\newacronym{mMIMO}{mMIMO}{massive multiple-input multiple-output}
\newacronym{SINR}{SINR}{signal-to-interference-noise ratio}
\newacronym{SNR}{SNR}{signal-to-noise ratio}
\newacronym{RSSI}{RSSI}{received signal strength indicator}
\newacronym{PZF}{PZF}{phase zero forcing}
\newacronym{PSO}{PSO}{particle swarm optimization}
\newacronym{ZF}{ZF}{zero forcing}
\newacronym{O-FDP}{O-FDP}{optimal fully digital precoder}
\newacronym{JT}{JT}{joint transmission}
\newacronym{CU}{CU}{central unit}
\newacronym{MSE}{MSE}{mean square error}
\newacronym{CEL}{CEL}{cross entropy loss}
\newacronym{CB}{CB}{conjugate beamforming}
\newacronym{NC}{NC}{network controller}
\newacronym{CoMP}{CoMP}{coordinated multi point}
\newacronym{CF-mMIMO}{CF-mMIMO}{cell-free massive MIMO}
\newacronym{CF-HBF}{CF-HBF}{cell-free hybrid beamforming}
\newacronym{CF-BF}{CF-BF}{cell-free beamforming}
\newacronym{MLDG}{MLDG}{Meta-Learning Domain Generalization}
\newacronym{MAML}{MAML}{Model Agnostic Meta-Learning}
\newacronym{WSR}{WSR}{Weighted Sum Rate}

\newif\ifDeepMIMOModel
\DeepMIMOModeltrue

\newif\ifSimpleNParamEq
\SimpleNParamEqtrue

\usepackage{lettrine}
\usepackage[ruled,vlined,linesnumbered]{algorithm2e}
\usepackage{multirow}
\usepackage{longtable}
\usepackage[table,xcdraw]{xcolor}
\usepackage{array} 
\makeatletter
\let\oldlt\longtable
\let\endoldlt\endlongtable
\def\longtable{\@ifnextchar[\longtable@i \longtable@ii}
\def\longtable@i[#1]{\begin{figure}[t]
\onecolumn
\begin{minipage}{0.5\textwidth}
\oldlt[#1]
}
\def\longtable@ii{\begin{figure}[t]
\onecolumn
\begin{minipage}{0.5\textwidth}
\oldlt
}
\def\endlongtable{\endoldlt
\end{minipage}
\twocolumn
\end{figure}}
\makeatother
\usepackage{ifpdf}
\ifCLASSINFOpdf
  \usepackage[pdftex]{graphicx}
  \usepackage{graphicx,epstopdf}
  \graphicspath{{../pdf/}{../jpeg/}}
  \DeclareGraphicsExtensions{.pdf,.jpeg,.png,.eps}
\else
  \usepackage[dvips]{graphicx}
  \usepackage{graphicx,epstopdf}
  \graphicspath{{../eps/}}
  \DeclareGraphicsExtensions{}
\fi
\usepackage{tikz}
\usetikzlibrary{decorations.pathreplacing,calc}

\usepackage{cite}
\usepackage{amsthm}
\usepackage{steinmetz}
\usepackage{amssymb}
\usepackage{balance}
\usepackage{eqparbox}
\usepackage{multirow}
\usepackage{bbm}
\usepackage{float}
\SetAlFnt{\small}

\usepackage{amsmath}

\usepackage{mathtools, nccmath}

\usepackage{booktabs, siunitx}
\usepackage{dcolumn}
\newcolumntype{d}[1]{D{.}{.}{#1}}
\usepackage[scaled]{DejaVuSansMono}
\usepackage[T1]{fontenc}

\usepackage{color}
\usepackage{relsize}
\usepackage{mathtools}

\newcommand{\bs}[1]{\boldsymbol{#1}}

\newcommand{\mb}[1]{\mathbf{#1}}

\usepackage{mathtools}

\SetKwInput{KwInput}{Input}                % Set the Input
\SetKwInput{KwOutput}{Output}
\SetKwInput{KwOutputr}{Output Regression}              % set the Output
\SetKwInput{KwOutputc}{Output Classification}              % set the Output

\newcommand{\bseq}{\begin{subequations}}
\newcommand{\eseq}{\end{subequations}}
\newcommand{\baln}{\begin{align}}
\newcommand{\ealn}{\end{align}}
\newcommand{\balnd}{\begin{aligned}}
\newcommand{\ealnd}{\end{aligned}}
\newcommand{\beq}{\begin{equation}}
\newcommand{\eeq}{\end{equation}}
\newcommand{\beqn}{\begin{eqnarray}}
\newcommand{\eeqn}{\end{eqnarray}}
\newcommand{\beqno}{\begin{eqnarray*}}
\newcommand{\eeqno}{\end{eqnarray*}}
\newcommand{\bma}{\begin{displaymath}}
\newcommand{\ema}{\end{displaymath}}
\newcommand{\bnu}{\begin{enumerate}}
\newcommand{\enu}{\end{enumerate}}
\newcommand{\bce}{\begin{center}}
\newcommand{\ece}{\end{center}}
\newcommand{\btb}{\begin{tabular}}
\newcommand{\etb}{\end{tabular}}
\newcommand{\ba}{\begin{array}}
\newcommand{\ea}{\end{array}}
\usepackage{footnote}
\makesavenoteenv{tabular}
\makesavenoteenv{table}
\makeatletter 
\newcommand\semiHuge{\@setfontsize\semiHuge{21.1}{27.38}}
\makeatother
\begin{document}
\title{Compression of Site-Specific Deep Neural Networks for Massive MIMO Precoding}
\author{\IEEEauthorblockN{Ghazal~Kasalaee, Ali~Hasanzadeh~Karkan, Jean-François~Frigon, and François~Leduc-Primeau}
\IEEEauthorblockA{Department of Electrical Engineering, Polytechnique Montréal, Montréal, QC H3C 3A7, Canada\\
Emails: \{ghazal.kasalaee, ali.hasanzadeh-karkan, j-f.frigon, francois.leduc-primeau\}@polymtl.ca
    }
}

\maketitle
\IEEEpubidadjcol

\begin{abstract}
The deployment of deep learning (DL) models for precoding in massive multiple-input multiple-output (mMIMO) systems is often constrained by high computational demands and energy consumption. In this paper, we investigate the compute energy efficiency of mMIMO precoders using DL-based approaches, comparing them to conventional methods such as zero forcing and weighted minimum mean square error (WMMSE). Our energy consumption model accounts for both memory access and calculation energy within DL accelerators. We propose a framework that incorporates mixed-precision quantization-aware training and neural architecture search to reduce energy usage without compromising accuracy. Using a ray-tracing dataset covering various base station sites, we analyze how site-specific conditions affect the energy efficiency of compressed models. Our results show that deep neural network compression generates precoders with up to 35 times higher energy efficiency than WMMSE at equal performance, depending on the scenario and the desired rate. These results establish a foundation and a benchmark for the development of energy-efficient DL-based mMIMO precoders.
\end{abstract}

% \begin{IEEEkeywords}
% \end{IEEEkeywords}
% \IEEEpeerreviewmaketitle

\section{Introduction} \label{Sec:Intro}
The deployment of \gls{DL} models in \gls{mMIMO} systems has shown great potential to advance wireless communication, particularly to optimize beamforming strategies. \Gls{DNN}s can dynamically adjust the beamforming to achieve better performance than traditional methods. However, the large size and overparameterization of these networks present challenges for practical deployment, especially on resource-constrained edge devices where memory, energy consumption, and processing latency are critically limited.

Designing energy-efficient communication systems is critical for the realization of \gls{mMIMO} systems. However, finding a precoding solution that maximizes energy efficiency under low-resolution quantizer constraints remains a significant challenge\cite{choi2022energy}. Quantized \glspl{CNN} are a popular choice for reducing hardware complexity. In \glspl{CNN}, quantization can be applied to inputs, weights, and activations~\cite{9150443}. Typically, weights and activations are represented using 32-bit or 64-bit floating-point formats, but they can be quantized to lower-precision fixed-point representations, such as 2-bit or even 1-bit precision, to further reduce computational and memory requirements. Lowering the precision leads to a smaller model size; however, it can also reduce accuracy, so choosing the number of precision bits is a non-trivial task.

These constraints hinder the direct application of DNNs in real-world mMIMO systems.
% , particularly when compared to more computational methods like Zero Forcing (ZF)~\cite{wiesel2008zero} and the \gls{WMMSE} algorithm~\cite{shi2023robust}. 
The celebrated \gls{WMMSE} algorithm proposed in~\cite{shi2011iteratively} is based on the equivalence between the gls{SINR} and mean square error, which is then solved using the block coordinate descent method. The WMMSE algorithm provides state-of-the-art performance and is widely used as a benchmark in the literature. However, existing iterative algorithms like WMMSE face challenges due to their computational complexity, which grows rapidly with the number of antennas and the network size. This increase in complexity leads to higher latency, making such algorithms less suitable for real-time applications or large-scale systems where rapid decision-making is critical.

To address these resource limitations, quantization techniques have emerged as an effective solution. Quantization reduces the complexity of neural networks by lowering the bit-width of model weights and activations, which decreases memory usage and energy consumption. Common approaches such as \gls{PTQ} and \gls{QAT} apply uniform precision across the entire network to optimize performance on resource-constrained platforms~\cite{720541, DBLP:journals/corr/HanMD15}. However, these methods often fail to account for the varying sensitivity of different network layers to quantization, leading to inefficiencies and potential performance loss when aggressive quantization is applied uniformly.

In contrast, \gls{MPQ} provides a more flexible approach by allowing different layers to use varying bit-widths, enabling a better balance between accuracy and energy efficiency~\cite{DBLP:journals/corr/abs-1805-06085}. By leveraging search algorithms such as Exhaustive Search, mixed-precision quantization can optimize bit-width allocation for specific layers, leading to significant reductions in energy consumption while maintaining high performance. This approach is particularly advantageous for complex models like \gls{DNN}s, which are widely used in energy-efficient tasks such as real-time signal processing in mMIMO systems.

This paper proposes a mixed-precision quantization framework, augmented with Neural Architecture Search (NAS)\cite{elsken2019neural}, to optimize energy consumption in DL models used for mMIMO beamforming. Our approach employs a streamlined neural architecture with a single CNN layer followed by three fully connected layers, designed to balance computational complexity and \gls{SINR} performance. Additionally, by adapting our quantization strategy to site-specific conditions using ray-tracing datasets, we further enhance the model's energy efficiency without sacrificing accuracy or SINR performance.

In this paper, we address the challenge of enhancing energy efficiency in \gls{DL}-based beamforming for \gls{mMIMO} systems. We propose a mixed-precision quantization-aware framework that significantly reduces energy consumption compared to standard deep learning methods while maintaining competitive sum rate performance. This approach achieves results comparable to traditional techniques such as \gls{ZF} and \gls{WMMSE}, with a strong emphasis on energy efficiency.
Additionally, we introduce site-specific model compression, a scalable method designed to adapt to unique site conditions. By leveraging site-specific ray-tracing datasets, this approach enables adaptive quantization that accounts for environmental factors, enhancing energy efficiency without compromising accuracy. 
This work establishes a solid foundation for advancing energy-efficient deep learning-based beamforming, offering a compelling alternative to conventional methods like ZF and WMMSE, particularly in scenarios where energy consumption is a critical concern.

\section{System model} \label{Sec:Baseline}
This work considers a multi-user \gls{mMIMO} system where a \gls{BS} with $N_{\sf{T}}$ transmit antennas serves $N_{\sf{U}}$ single-antenna users simultaneously. Let $x_u$ represent the transmitted symbol for each user. The received signal at the $u^{th}$ user can be represented as 
\begin{equation}\label{eq:signal_recived}
    \mathbf{y}_u =  \mathbf{h}_{u}^{\dagger} \sum_{\forall u}  \mathbf{w}_{u} x_u + \bs{\eta} \,,
\end{equation}
in which $\mathbf{h}_u \in \mathbb{C}^{N_{\sf{T}} \times 1}$ is the wireless channel vector between the \Gls{BS} and $u^{th}$ user; the term $\bs{\eta} \sim \mathcal{CN}(0, \sigma^2)$ denotes complex symmetric Gaussian noise with zero mean and a variance of $\sigma^2$; the downlink transmit \gls{FDP} vector is denoted as $\mathbf{W} = \left [ \mathbf{w}_1, \hdots, \mathbf{x}_u, \hdots, \mathbf{w}_{N_{\sf{U}}} \right ] \in \mathbb{C}^{N_{\sf{T}} \times N_{\sf{U}}}$.
The corresponding \gls{SINR} for user $u$ is formulated as
\begin{equation}
    \text{SINR}(\mb{w}_{u}) = \frac{ \big|\mb{h}^{\dagger}_{u} \mb{w}_{u} \big|^2}{\sum_{j \neq u} \big|\mb{h}^{\dagger}_{u} \mb{w}_{j} \big|^2 + \sigma^2} \,.
\end{equation}
The goal is to find such a precoding $\mathbf{W}$ that maximizes the throughput under the maximal transmit power $P_{\text{max}}$ constraint. Thus, the downlink sum rate maximization problem can be formulated as
\begin{align}\label{eq:sum rate-optimization}
    & \underset{\mb{W}}{\max}~ R(\mb{W}) \,, \\
    \text{s.t.}  &\sum_{\forall u} \mb{w}_{u}^{\dagger}  \mb{w}_{u} \leq P_{\sf{max}}\,,
\end{align}
where for a precoding matrix $\mathbf{W}$, the sum rate is
\begin{equation}\label{eq:sum rate}
    R(\mb{W}) = \sum_{\forall u} \text{log}_2 \Bigl(  1+ \text{SINR}(\mb{w}_{u}) \Bigr).
\end{equation}

\subsection{Problem Definition}

Beamforming is crucial in wireless communication, particularly at mmWave frequencies, where directional signal transmission is essential due to the limitations of omnidirectional antennas. \Glspl{DNN} have shown promise in maximizing the sum rate, but despite their enhanced performance, \glspl{DNN} come with high computational complexity and substantial energy requirements, posing challenges for deployment on resource-limited devices like edge platforms.

To tackle the complexities and energy demands associated with \gls{DNN}-based precoders in mMIMO systems, we focus on developing a method that preserves the high throughput of \glspl{DNN} while making them suitable for resource-constrained environments. This requires overcoming inefficiencies caused by uniformly applied quantization, which overlooks the varying sensitivity of different network layers to precision reduction.

In this work, we introduce a novel framework that combines mixed-precision quantization-aware training with \gls{NAS}. Our method is designed to significantly lower the energy consumption of \glspl{DNN} in precoder design while sustaining high throughput for site-specific \glspl{BS}. Extensive experiments validate that our approach achieves superior energy efficiency compared to existing \gls{DL}-based and conventional beamforming techniques.

\section{Methodology} \label{Sec:Proposed}
\begin{figure}[!t]
\includegraphics[width=\linewidth]{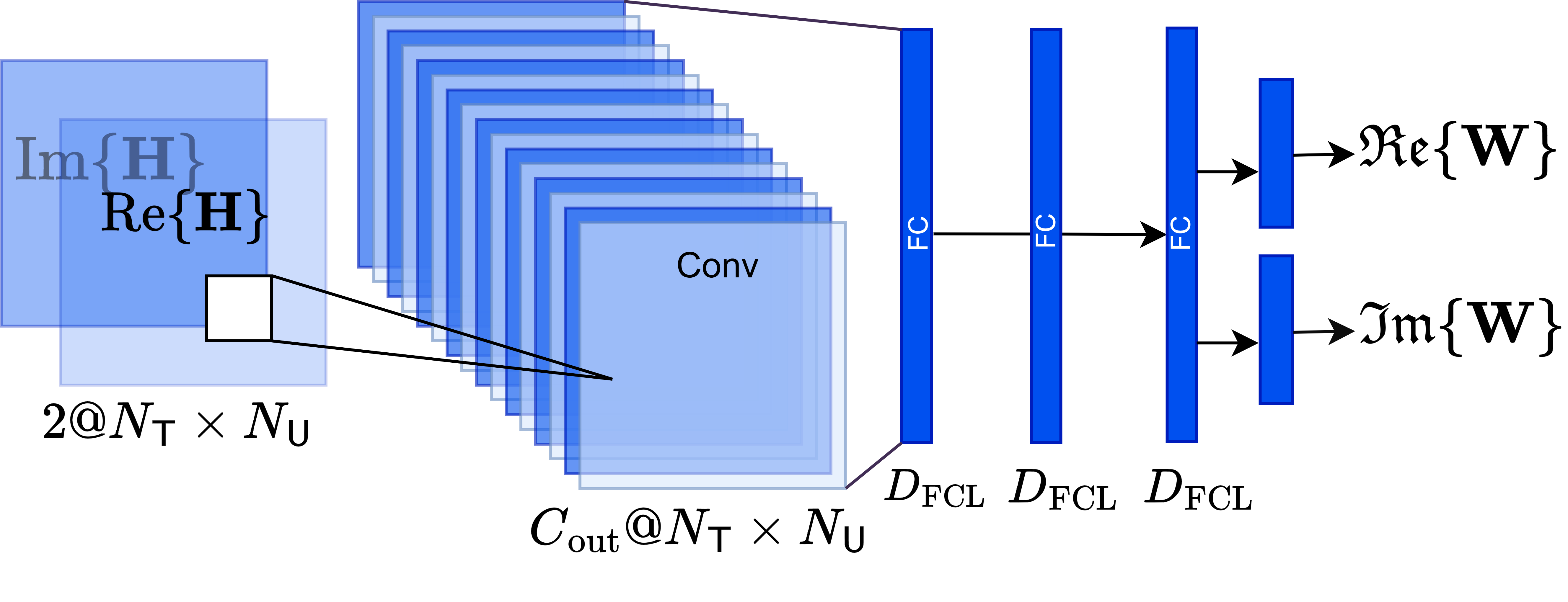}
\caption{DNN architecture template.}
\label{fig:DNN}
\end{figure}
\subsection{Model Architecture and NAS}
The proposed \gls{DNN} architecture for digital beamforming, illustrated in Figure \ref{fig:DNN}, is specifically designed to address the demands of \gls{mMIMO} systems. The architecture gets the real and imaginary parts of the channel state information through two separate channels as input and predicts the precoding matrix. The network comprises a single \gls{CNN} layer with the kernel size of 3 for feature extraction, followed by three \glspl{FCL} to map the extracted features into the real and imaginary components of the digital precoder. To enhance model generalization and stability, the architecture incorporates ReLU activation functions, 5\% dropout to mitigate overfitting, and batch normalization to accelerate convergence and improve robustness during training.
Our approach optimizes quantization by exploring a search space of bit-width configurations for the weights across the layers of the \glspl{DNN}. We vary the precision levels for each layer—[CONV, FCL1, FCL2, FCL3]—with multiple bit-width choices, allowing for a large number of possible permutations. Additionally, we consider different model architectures by varying the output channel size \(C_{\text{out}}\) of the convolutional layer and the size \(D_{\text{FCL}}\) of \glspl{FCL}. This results in a comprehensive search space combining quantization and architectural configurations. We use exhaustive search to explore all possible combinations, ensuring that optimal configurations are identified without overlooking any potential solutions. The small model size and fast training times make this exhaustive search feasible. Ultimately, this method helps to balance energy efficiency and sum rate performance, which is essential for energy-constrained \gls{mMIMO} systems where optimizing both performance and resource usage is critical.

\subsection{Quantization Methods}

This study examines several quantization techniques to enhance energy efficiency and maintain performance in \glspl{DNN} for \gls{mMIMO} systems. The main objective is to minimize energy consumption and memory demands while preserving an acceptable sum rate. We explore \gls{QAT} using the \gls{LSQ} method \cite{Esser2020LEARNED} and \gls{MPQ} through an exhaustive search. These quantization methods are then combined with \gls{NAS} to find an optimal balance between energy efficiency and performance.

\subsubsection{\gls{PTQ}}
In preliminary experiments, we also considered \gls{PTQ} as a simple way to reduce model complexity.
This method applies fixed-point quantization to a pre-trained floating-point (FP32) model without requiring retraining. 
However, in our PTQ experiments, sum rate performance was highly degraded at low resolutions (4 bits or less), and therefore we focus in this paper on the \gls{QAT} approach to investigate the best possible model compression.

\subsubsection{\gls{QAT}}
To address the limitations of \gls{PTQ}, \gls{QAT} fine-tunes a pre-trained model within quantization constraints, enabling the network to adapt to lower precision with minimal performance loss. We use \gls{LSQ}\cite{Esser2020LEARNED}, which dynamically adjusts quantization step sizes during training. Unlike static step quantization, \gls{LSQ} optimizes these step sizes to mitigate quantization error, especially in low-bit scenarios. This is achieved through the Straight Through Estimator \cite{DBLP:journals/corr/abs-2109-05472}, which supports gradient-based optimization for non-differentiable quantization functions. The adaptive nature of \gls{LSQ} provides significant gains in both sum rate and energy efficiency under low-bit quantization constraints.

\subsubsection{\gls{MPQ} with \Gls{NAS}}
\gls{MPQ} assigns different bit-widths to each network layer, offering a flexible trade-off between energy efficiency and accuracy. We employ exhaustive search to evaluate quantization combinations [2,\,4,\,8,\,and\,16\,bits] across four layers, calculated as \(4^4 \)\,$=$\,\(256\).
and various model architectures by varying \( C_{\text{out}} \in \{8,\,16,\,32,\,64\} \) and \( D_{\text{FCL}}\)\,$\in$\,\(\{512, 1024\} \) across the \Gls{CNN} and \glspl{FCL} (see Fig.~\ref{fig:DNN}) to minimize energy consumption while preserving sum rate performance \cite{DBLP:journals/corr/abs-2003-07577}.

\subsection{Training Method}
All models are trained using a self-supervised approach \cite{10624768}, with sum-rate maximization as the objective function. This method ensures that the model learns to predict effective precoding vectors directly, without the need for labeled data. The loss function is defined as 
\begin{align}
    \mathcal{L} = - R(\mb{W})\, ,
\end{align}
and models are trained with a fixed learning rate of $10^{-3}$ and a batch size of 1000.
Additionally, we average the results from four training runs with distinct initialization seeds to control for variations in the training process.

\subsection{Energy Model for Quantized Neural Networks}\label{sec:energy_dnn}
In order to evaluate and compare the energy consumption of a \gls{DNN} solution, we consider a simple but realistic model of the energy consumed to compute the \gls{DNN} output.
We base our model on the one proposed in \cite{moons:2017a}, which considers the energy required for memory accesses and computations, while taking into account the impact of the bit width of model weights and activations.

The energy consumption of the DNN is thus decomposed into three components: computation energy (\(E_C\)), weight transfer energy (\(E_W\)), and activation transfer energy (\(E_A\)). 
The total energy consumption is then given by
\begin{equation}\label{eq:total-energy}
    E_{\mathrm{DNN}} = E_C + E_W + E_A \,.
\end{equation}
The computation energy is based on counting the number of \gls{MAC} operations required for the linear portion of each layer and the number of biasing, batch normalization, and activation function computations. To simplify the model, all these operations are attributed an energy $E_{\text{MAC}}$.
The computation energy $E_C$ is thus given by
\begin{equation}\label{eq:computation-energy}
    E_C = E_{\text{MAC}} \cdot (N_c + 3 \cdot N_a)\,,
\end{equation}
where $N_c$ is the total number of MAC operations required by the model, $N_a$ is the total number of activations, and the factor $3$ arises since one biasing, one normalization, and one activation must be computed for each activation output.

For consistency with the baseline energy model that will be presented next, we model $E_\text{MAC}$ in terms of the bit width $Q$ as
\begin{equation}\label{eq:mac-energy}
    E_{\text{MAC}} = \alpha \cdot \left( \frac{Q}{16} \right)^{\beta}\,,
\end{equation}
while choosing $\alpha$\,$=$\,$0.86$ and $\beta$\,$=$\,$1.9$ to fit the energy measurements reported in \cite{6757323}.

The energy associated with transferring weights, \(E_W\), is expressed as
\begin{equation}\label{eq:weight-energy}
    E_W = E_M \cdot N_w + E_L \cdot \frac{N_c}{\sqrt{p}},
\end{equation}
where \(N_w\) is the total number of weights, \(E_M\) and \(E_L\) represent the energy costs of accessing the main on-chip memory and local buffers, respectively, and $p$ is the number of parallel execution units.
We use $E_M=2 E_L=2 E_\mathrm{MAC}$ and  \(p =
64 \big(\frac{Q}{16}\big)\).
It is assumed that the entire DNN model fits on-chip and no accesses to external memory are needed.

Finally, the energy for transferring activations, \(E_A\), is given by
\begin{equation}\label{eq:activation-energy}
    E_A = 2 \cdot E_M \cdot N_a + E_L \cdot \frac{N_c}{\sqrt{p}}\,.
\end{equation}

%This energy model captures the interplay between computational and off-memory access energy costs, offering a framework for analyzing and optimizing \glspl{QNN} under varying quantization levels and hardware configurations.

\subsection{Energy Model for Baseline Methods}\label{sec:energy_baselines}
We compare the energy consumption of the proposed quantized DNN models to conventional algorithms, specifically the \gls{WMMSE} and \gls{ZF} methods. The \gls{WMMSE} algorithm, known for its high computational complexity due to iterative processing and matrix inversions, consumes significantly more energy than the \gls{ZF} approach. 
Since these energy models act as a baseline, we adopt a conservative (lower bound) approach by accounting only for the multiplications when estimating compute energy. For memory energy estimation, we consider only the local buffer accesses for the operands.
The total number of real multiplications required for $I$ iterations of \gls{WMMSE} is given by
\begin{align}
 N_c = I\bigg(\frac{8}{3} N_{\sf{T}}^3N_{\sf{U}} &+ 4N_{\sf{T}}^2N_{\sf{U}} 
 + 4N_{\sf{T}}(4N_{\sf{U}}^2+2N_{\sf{U}})\notag\\
 &+ 4N_{\sf{U}}^2
 +\frac{56}{3}N_{\sf{U}}\bigg)  \, ,
\end{align}
% \begin{align*}
%     I\bigg(\frac{2}{3} N_{\sf{T}}^3 N_{\sf{U}} 
%     + N_{\sf{T}}^2 N_{\sf{U}} (N_{\sf{U}} + 2)
%     + N_{\sf{T}}N_{\sf{U}}(N_{\sf{U}}-1) 
%     - \frac{1}{3}N_{\sf{U}}\bigg)\, 
% \end{align*}
while the required number of real multiplications for ZF is
\begin{equation}
    N_c = 8 N_{\sf{U}}^2 N_{\sf{T}} + \frac{8}{3} N_{\sf{U}}^3 \, .
\end{equation}
% \begin{align*}
%     2N_{\sf{U}}^2 (2 N_{\sf{T}} - 1) + \frac{4}{3} N_{\sf{U}}^3.
% \end{align*}
The energy model for WMMSE and ZF can then be expressed as
\begin{equation}
    E_{\mathrm{B}} = E_{MAC} \cdot N_c + E_L\cdot\frac{N_c}{\sqrt{p}} \, .
\end{equation}
% (not the right place for a discussion) These models highlight a trade-off between computational complexity and energy efficiency: \gls{WMMSE} offers higher precision but at significantly higher energy costs, while \gls{ZF} provides a more energy-efficient alternative.

\section{Numerical Results} \label{Sec:Simulation}
\subsection{Dataset Definition}
A custom dataset was generated to accurately reflect the channel characteristics pertinent to beamforming in mMIMO systems. MATLAB’s Ray-Tracing toolbox simulated both \gls{LOS} and \gls{NLOS} conditions. The simulations positioned a \gls{BS} within the Montreal region, utilizing environmental data from OpenStreetMap \cite{OpenStreetMap} for realism. The base station employed a uniform planar array antenna with 8x8 elements, spaced at half-wavelength, operating at a frequency of 2\,GHz. The transmitter was set at a height of 20\,m and powered at 20\,W, assuming a system loss of 10\,dB. Users were placed in circular patterns around the base station at distances ranging from 50 to 350\,m and at 10-degree intervals. This configuration captured a wide variety of deployment scenarios and channel conditions. The ray-tracing simulations considered up to 10 reflections but excluded diffraction effects, focusing on signal reflections from buildings and terrain to emulate multipath propagation in urban environments accurately.

\subsection{Energy Consumption Examples}
We first present some examples of the energy consumed by the different approaches, as per the model presented in Sections~\ref{sec:energy_dnn} and \ref{sec:energy_baselines}.
We consider the ``UdeM-NLOS'' scenario.
In Table~\ref{table:energy_comparison}, we report the energy for the ``default'' variants of the method, that is, for \gls{WMMSE}, we set the stopping criterion to $10^{-5}$ to have near-optimal performance, and for the DNN, we use $C_{\text{out}}$\,$=$\,$64$ and $D_{\text{FCL}}$\,$=$\,$1024$ with the maximum weight resolution of 16 bits.
We see that the DNN consumes significantly less than WMMSE at the cost of a slight degradation in sum rate (on this scenario). ZF, on the other hand, is much less complex, but does not provide competitive performance in \gls{NLOS} conditions, or at low \gls{SNR}.
As a result, ZF is unlikely to be a favored solution in practice, since it results in significant under-utilization of the BS resources.

\begin{table}[t]
    \centering
    \caption{Energy Comparison of the Default Variants on ``U\lowercase{de}M-NLOS''  ($N_{\sf{T}}$\,$=$\,$64$,~$N_{\sf{U}}$\,$=$\,$4$,~SNR\,$=$\,15\,dB).}
    \begin{tabular}{ccc}
        \toprule
        \textbf{Method} & \textbf{Energy ($\mu J$)} & \textbf{Sum Rate} (bit/s/Hz) \\
        \midrule
        WMMSE ($I$\,$=$\,$92.8$) & 296 & 19.2\phantom{$~\pm~0.007$}\\
        \textbf{Default DNN} & \textbf{54.5} & \textbf{18.9}$~\pm~0.007$\\
        Zero-Forcing (ZF) & 0.008  & 15.3\phantom{$~\pm~0.007$}\\
        \bottomrule
    \end{tabular}
    \label{table:energy_comparison}
\end{table}

\begin{table}[t]
    \centering
    \caption{DNN Energy Consumption with Uniform Quantization 
    \\for $C_{\text{out}}$\,$=$\,$64$, $D_{\text{FCL}}$\,$=$\,$1024$ 
    on ``U\lowercase{de}M-NLOS''\\
    ($N_{\sf{T}} $\,$=$\,$ 64$, $N_{\sf{U}}$\,$=$\,$4$, SNR\,$=$\,15\,dB)}
    \resizebox{\columnwidth}{!}{
    \begin{tabular}{ccc}
        \toprule
        \textbf{Quantization Configs}
        & \textbf{Energy ($\mu J$)}& \textbf{Sum Rate} (bit/s/Hz)\\
        \midrule
        $[16,\,16,\,16,\,16]$& 54.5 & 18.9$~\pm~0.007$\\
        $[8,\,8,\,8,\,8]$ & 17.5 & 18.6$~\pm~0.012$ \\
        $[4,\,4,\,4,\,4]$ & 7.4 & 17.8$~\pm~0.014$\\
        $[2,\,2,\,2,\,2]$ & 4.6 & 17.1$~\pm~0.032$\\
        \bottomrule
    \end{tabular}
    }
    \label{tab:energy_consumption}
\end{table}

Next, to illustrate the impact of quantization, 
Table~\ref{tab:energy_consumption} lists the energy consumption of the DNN for various uniform bit-width configurations, for the same $C_{\text{out}}$ and $D_{\text{FCL}}$ as in Table~\ref{table:energy_comparison}. We see that lowering the weight resolution leads to substantial energy savings but at the cost of a moderate decrease in performance.

\begin{figure}[!t]
    \centering
    \includegraphics[width=\columnwidth]{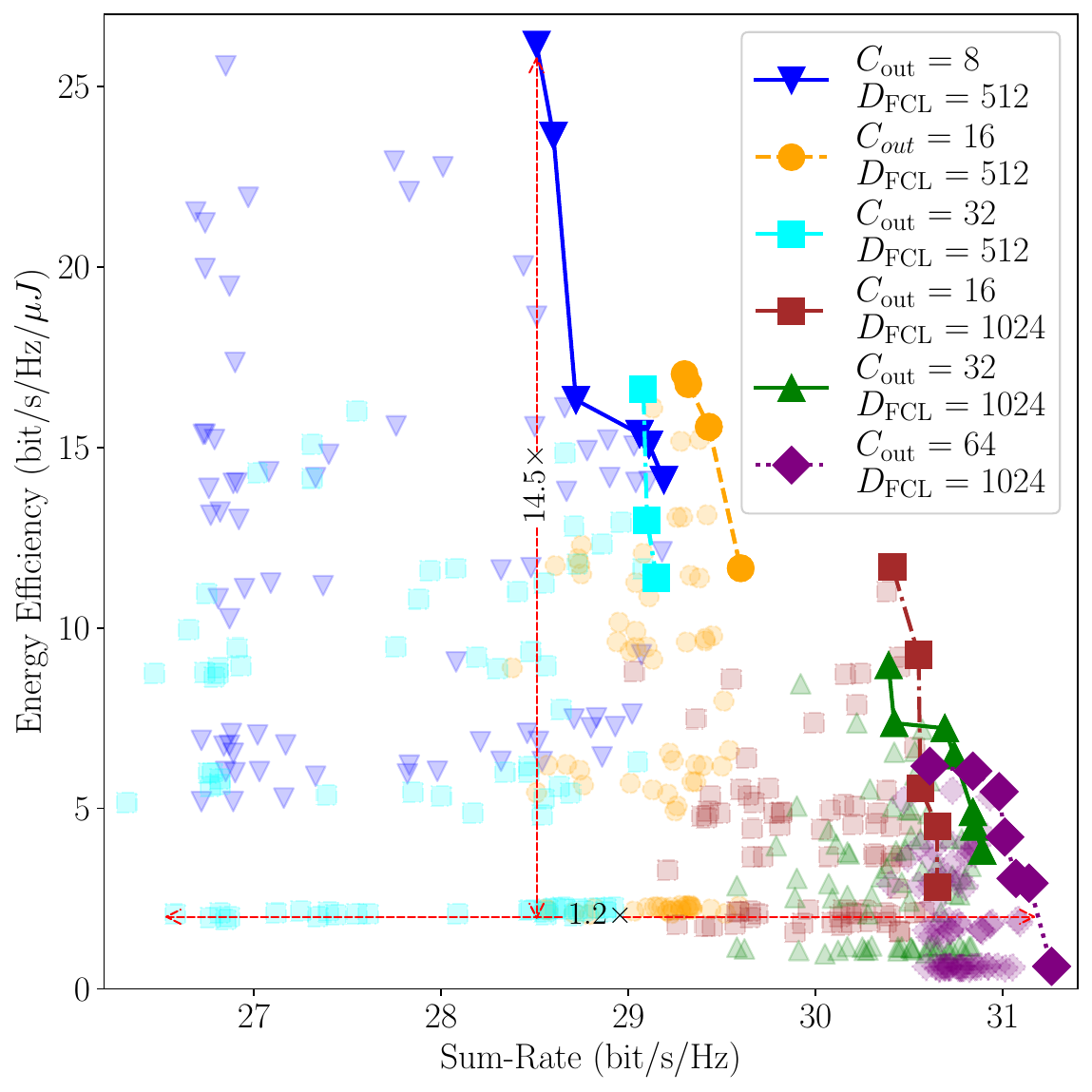}
    \caption{Trade-off between energy efficiency and sum rate for \glspl{CNN} with varying $C_{\text{out}}$, $D_{\text{FCL}}$, and MPQ bit widths, on the ``UdeM-LOS'' scenario ($N_{\sf{T}}$\,$=$\,$64$, $N_{\sf{U}}$\,$=$\,$4$, average SNR\,$=$\,29\,dB). All the model configurations that were evaluated are shown, while the curves provide the Pareto front associated with each architecture configuration.}
    \label{fig:Neural Architecture Search (NAS)}
    \vspace{-10pt}
\end{figure}

\subsection{NAS Results}

Figure \ref{fig:Neural Architecture Search (NAS)} highlights the trade-offs between computational energy efficiency (bits/s/Hz/$\mu$J) and sum rate (bits/s/Hz) across diverse \gls{DNN} configurations generated as described in Section~\ref{Sec:Proposed}, on the ``UdeM-LOS'' scenario.
Each curve shows the Pareto front corresponding to a particular architecture configuration, while each point in this curve uses a different quantization configuration.

A few trends can be mentioned among the Pareto-optimal results for each architecture.
Firstly, the first layer, CONV, is often kept at high precision, particularly in smaller models, to maintain performance while reducing energy consumption. Interestingly, no Pareto-optimal model uses uniform quantization across all layers. Moreover, even models that achieve the highest sum rates do not employ more than two layers at the highest precision. These results emphasize the importance of efficiently distributing bit precision across layers to optimize energy consumption.

The configuration yielding the highest energy efficiency is ($C_{out}$\,$=$\,$8$, $D_{FCL}$\,$=$\,$512$) with quantization [2,\,8,\,8,\,8], achieving 26.2 \,bits/s/Hz/$\mu$J at a moderate sum rate of 28.5\,bits/s/Hz. On the other hand, the configuration achieving the highest sum rate is ($C_{out}$\,$=$\,$64$, $D_{FCL}$\,$=$\,$ 1024$) with quantization [16,\,16,\,4,\,8], which reaches 31.3\,bits/s/Hz but does not use full precision, making it more interesting from an energy efficiency perspective. An alternative worth mentioning is the model with the second-best sum rate of 31.1\,bits/s/Hz, achieved with an architecture of ($C_{out}$\,$=$\,$64$, $D_{FCL}$\,$=$\,$1024$) and quantization [16,\,2,\,2,\,2], which uses nearly $5\times$ less energy than the highest sum-rate model.
We do observe compute energy efficiency decreasing rapidly near the highest sum-rate, but of course this could simply mean that switching to a larger and/or different DNN architecture would be preferable at that point.

To further illustrate the importance of of NAS and model compression in finding efficient DNN precoders, Figure~\ref{fig:Neural Architecture Search (NAS)} includes horizontal and vertical arrows that quantify the impact on performance of the design choices. The horizontal arrow measures the difference in sum rate between the worst and best configurations at equal energy efficiency, revealing a 20\% improvement through optimal model selection. Similarly, the vertical arrow shows the difference in energy efficiency at an equal sum rate, demonstrating a $14.5\times$ gain.

\begin{figure}[!t]
\centering
\includegraphics[width=\columnwidth]{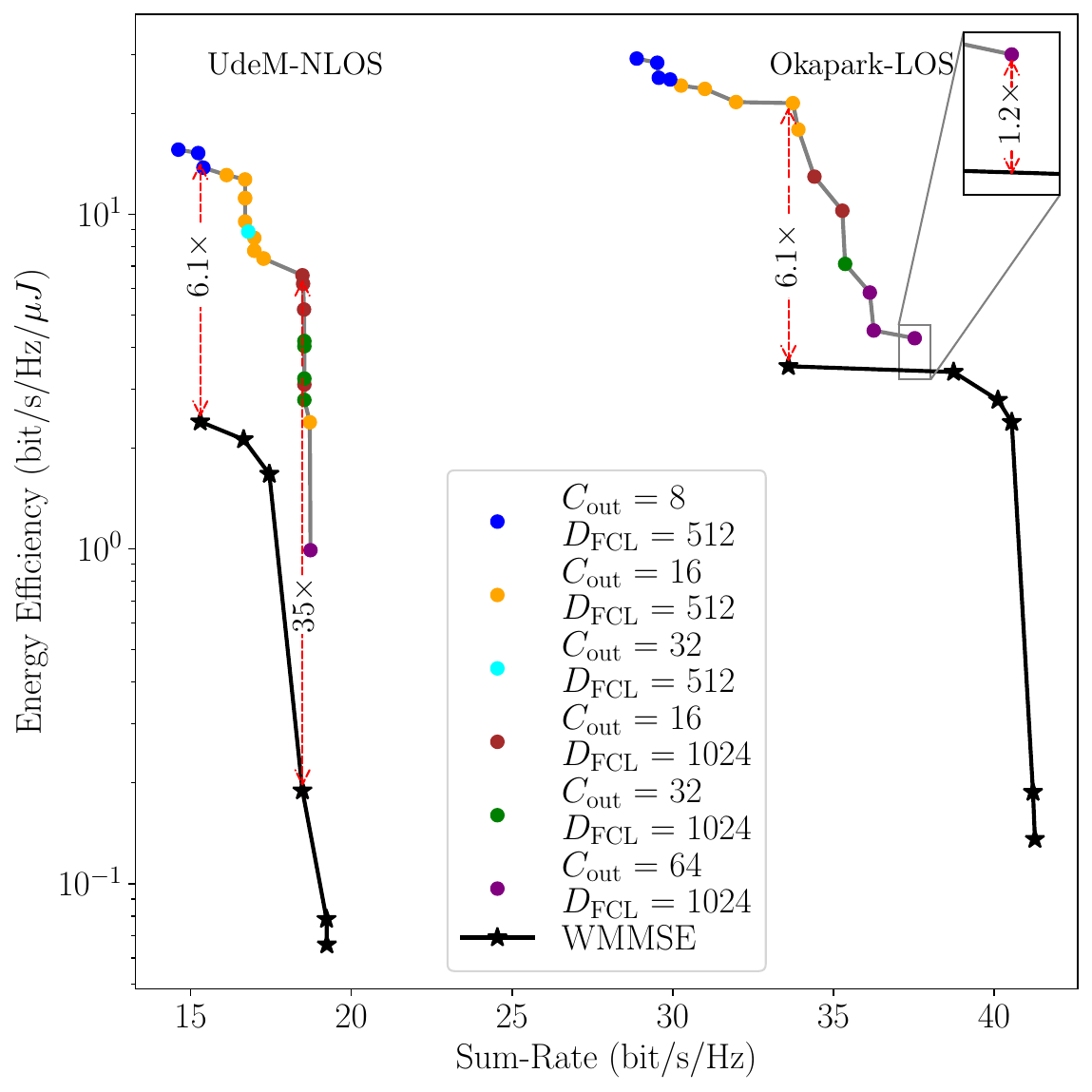}
\caption{Comparison of energy efficiency (bits/s/Hz/$\mu$J) and sum rate (bits/s/Hz) across two environments: UdeM-NLOS (average SNR\,$=$\,15\,dB) and Okapark-LOS (average SNR\,$=$\,28\,dB). The proposed method achieves a superior balance of energy efficiency and sum rate performance compared to WMMSE. Results are derived for models with varying ($C_{out}$) and ($D_{FCL}$).}
\vspace{-9pt}
\label{fig:LSQ Quantization}
\end{figure}
\subsection{Impact of Deployment Environment and Energy Efficiency Comparison}

Figure~\ref{fig:LSQ Quantization} compares energy efficiency (bits/s/Hz/$\mu$J) and sum-rate (bits/s/Hz) across two contrasting deployment scenarios: UdeM-NLOS, characterized by challenging multipath conditions, and Okapark-LOS, offering clear \gls{LOS} signal propagation. These two scenarios highlight the adaptability and effectiveness of the proposed quantized models.
For the DNN precoders, each curve shows the Pareto-optimal configurations across the entire search space, whereas for WMMSE, the trade-off between sum-rate and energy efficiency is varied by adjusting the stopping criterion.

In the UdeM-NLOS environment, sum rates are constrained between 15 and 20 bits/s/Hz due to severe signal attenuation and multipath effects. Despite these limitations, the quantized models achieve significant energy savings, with improvements of up to $35\times$ in energy efficiency compared to the WMMSE baseline, all while maintaining competitive sum rates.

In contrast, the Okapark-LOS environment, which benefits from clear signal paths, supports higher sum rates ranging from about 30 to 40 bits/s/Hz. 
Depending on the desired sum-rate, the DNN precoders can provide improvements in energy efficiency ranging from $6.1\times$ to $1.2\times$. However, with the DNN architecture template and training method considered in this paper, the DNN precoder is unable to achieve the highest sum-rate that can be provided by WMMSE.
%Here, the quantized models continue outperforming the WMMSE baseline, delivering a 6.1× improvement in energy efficiency while maintaining similar or superior sum-rate performance. The highest sum rate, achieved by the DNN for Okapark-LOS ($C_{out} = 64$, $D_{FCL} = 1024$), is roughly 1.3× higher, highlighting an intriguing balance of performance and precision.

These results emphasize the adaptability of the quantized models across diverse deployment environments, and their ability to achieve the same performance with less compute energy. Interestingly, the energy gains provided by DNNs appear to be larger in more difficult (low SNR, non line-of-sight) environments.

\section{Conclusion} \label{Sec:Conclusion}
We proposed a novel framework for finding efficient compressed DNN models for \gls{mMIMO} precoding.
By using quantization-aware training with \gls{LSQ} and \gls{MPQ}, while also searching through different DNN architecture sizes, we find DNN models with significantly lower energy consumption at equal performance. 
A variety of Pareto-optimal models were generated, showing the ability of DNN solutions to provide different tradeoffs between energy consumption and performance.
%Training different model sizes within NAS allowed us to adaptively balance energy efficiency with sum rate performance. 
Compared to \gls{WMMSE}, the obtained DNN models achieve superior energy efficiency across diverse deployment scenarios. This work demonstrates the promise of DNN solutions to obtain high-performance \gls{mMIMO} precoders with much improved energy efficiency.

\section*{Acknowledgement}
This work was supported by Ericsson - Global Artificial Intelligence Accelerator AI-Hub Canada in Montr\'{e}al and jointly funded by NSERC Alliance Grant 566589-21 (Ericsson, ECCC, Innov\'{E}\'{E}).

\bibliographystyle{IEEEtran}
% \bibliography{am_ger_eng,rubi_eng}
% \bibfont \footnotesize
\bibliography{0_main}

\end{document}